\documentclass[%
 reprint,
superscriptaddress,
 amsmath,amssymb,
 aps,
 prl,
]{revtex4-1}

\usepackage{graphicx}
\usepackage{dcolumn}
\usepackage{bm}
\usepackage[colorlinks= true, linkcolor=blue, citecolor=blue, urlcolor=blue]{hyperref}
\usepackage{lipsum}
\usepackage{bbold}
\usepackage{mathtools}
\usepackage[normalem]{ulem}

\def\bea{\begin{eqnarray}}
\def\eea{\end{eqnarray}}

\usepackage{xcolor}

\begin{document}


\title{Arrhenius law for interacting diffusive systems}

\author{Vishwajeet Kumar}
\affiliation{The Institute of Mathematical Sciences, CIT Campus, Taramani, Chennai 600113, India \& Homi Bhabha National Institute, Training School Complex, Anushakti Nagar, Mumbai 400094, India}
\author{Arnab Pal}
\email{arnabpal@imsc.res.in}
\affiliation{The Institute of Mathematical Sciences, CIT Campus, Taramani, Chennai 600113, India \&
Homi Bhabha National Institute, Training School Complex, Anushakti Nagar, Mumbai 400094, India}
\author{Ohad Shpielberg}
\email{ohads@sci.haifa.ac.il}
\affiliation{Department of Mathematics and Physics, University of Haifa at Oranim, Kiryat Tivon 3600600, Israel}
\affiliation{Haifa Research Center for Theoretical Physics and Astrophysics,
University of Haifa, Abba Khoushy Ave 199, Haifa 3498838, Israel}


\begin{abstract}
Finding the mean time it takes for a particle to escape from a meta-stable state due to thermal fluctuations is a fundamental problem in physics, chemistry and biology. For weak thermal noise, the mean escape time is captured by the Arrhenius law (AL). Despite its ubiquity in nature and wide applicability in practical engineering, the problem is typically limited to single particle physics. Finding a generalized form of the AL for interacting particles has eluded solution for a century. Here, we tackle this outstanding problem and generalize the AL to a class of interacting diffusive systems within the framework of the macroscopic fluctuation theory. The generalized AL is shown to conform a non-trivial yet elegant form that depends crucially on the particle density and inter-particle interactions. We demonstrate our results for the paradigmatic exclusion and inclusion processes  to underpin the key effects of repulsive and attractive interactions. Intriguingly, we show how to manipulate the mean escape time using not only temperature, but also the particle density.  
\end{abstract}

\pacs{Valid PACS appear here}
\maketitle

\emph{Introduction.---}
The celebrated Arrhenius law (AL) is a cornerstone in physics, chemistry and biology, capturing the activation time of a system from a meta-stable state. Thermally induced activation processes are ubiquitous in nature,  e.g. in chemical reactions, protein folding, gene expressions to name but a few. Although each variant has its unique and intriguing features, the universality of AL with regard to the activation barrier and the surrounding temperature nonetheless is remarkable. 

Often, the AL can be manifested within the Kramer's reaction rate theory \cite{hanggi1990reaction}. Usually there, one is interested in the time taken by an overdamped particle to escape from a trapping potential $U(x)$ while coupled to a thermal bath at temperature $T$ \cite{gardiner1985handbook}.  For the particle to escape the trap, it needs a fluctuation to grant it an excess energy $\Delta U$ -- the energy difference between the bottom of the potential to the escape point at the top. The AL states that for weak thermal fluctuations $D_0 = k_B T \ll \Delta U $, the inverse of the mean escape time is   
\begin{align}
 \Phi = \tau^{-1} _0 \rm{e}^{ -\Delta U/D_0}.   
 \label{Kramers-1}
\end{align} 
$\tau_0$ provides a microscopic time scale, which may be assessed due to arguments by Eyring \cite{bouchet2016generalisation,eyring1935activated}. Importantly, $\tau_0$ is non-universal as it depends on the shape of $U(x)$. The wide applicability of the AL can be attributed to the universal exponential decay $\rm{e}^{- \Delta U/D_0}$. It suggests that by varying the temperature, the experimentally accessible mean escape time allows to infer the activation energy $\Delta U$, independent of the non-universal prefactor $\tau_0$. To understand this better, imagine a diffusive process on a multidimensional, rugged energy landscape that can imitate chemical reactions in a network or protein conformational dynamics from unfolded to natively folded state via misfolded state. Presence of a variety of local minima surrounded by energy barriers $\Delta U \gg k_B T$  renders a natural separation of timescales in these systems -- fast fluctuations in the well followed by slow/rare fluctuations between the wells. In other words, the
enzyme fluctuates many times within a well (typical trajectories) before leaving it (rare trajectories).
This is a key assumption behind Eq.~\eqref{Kramers-1} in generic activation processes \cite{carmeli1983theory,Kramer-ntizan,moffitt2010methods,makarov2015single}.

Despite many years of study, discoveries are still being made around the AL and exciting applications continue to be found such as activation in the presence of viscoelastic medium \cite{ginot2022barrier,ferrer2021fluid} or escape dynamics of active particles \cite{militaru2021escape,woillez2020active},  temperature dependent activation energies \cite{peleg2012arrhenius,dyre2006colloquium}, multiple meta-stable states \cite{militaru2021escape} and experiments with colloids \cite{thorneywork2020direct,chupeau2020optimizing}. Yet, one frontier that remains surprisingly less explored is the \textit{validity of AL in many-body systems}. Indeed, even for two particles with short-range interactions, finding the activation time seems to be a formidable challenge. See \cite{langer1969statistical} which sketches out the challenges for formulating a theory for interacting systems. In addition to the existing slow and fast time scales, the `nature' of interaction also sets another time-scale in the problem. Consequently, the universal exponential decay in AL may no longer hold true due to the complex interplay of different agents and their interactions \cite{stirnemann2012communication,NonArrhenius_IslandDensity,junior2019characterization,hennig2015cooperative}. This is illustrated in the breakdown of the AL for active particles \cite{woillez2019activated}, in the case of infinite range interactions \cite{mukamel2005breaking,saadat2023lifetime} as well as in low temperature glassy dynamics \cite{dyre2006colloquium}.

\begin{figure*}[t!]
\begin{center}
\includegraphics[width=14cm]{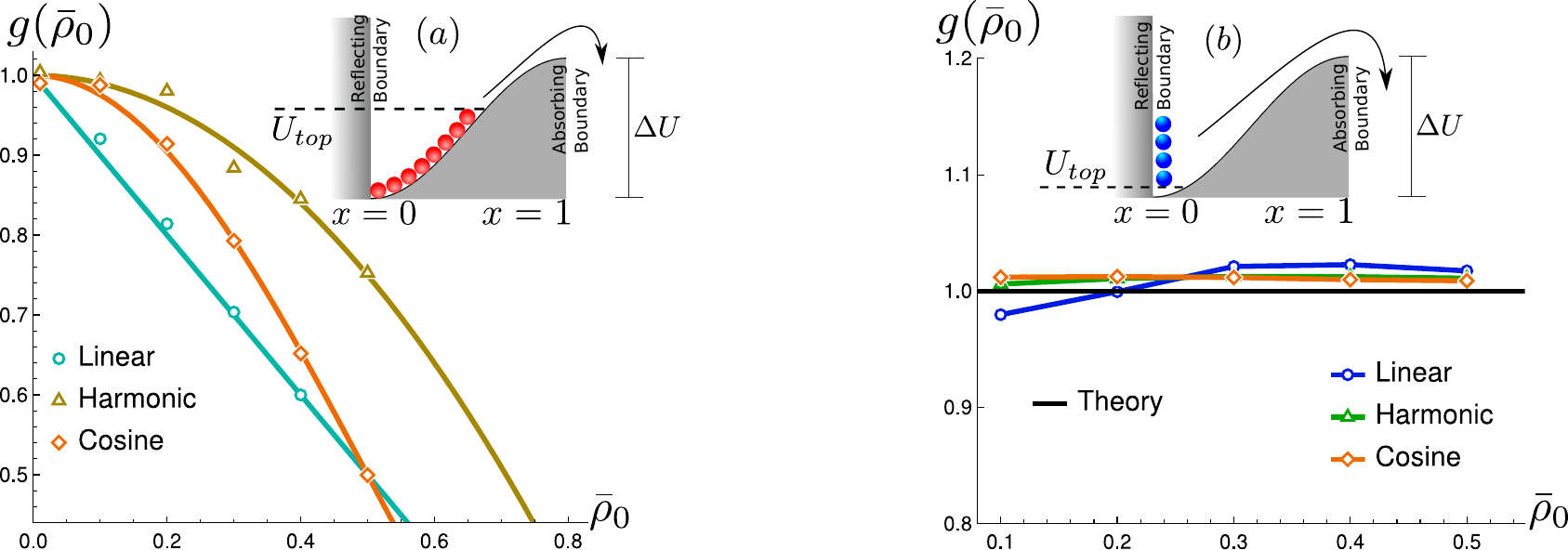}
\caption{The generalized AL for interacting diffusive systems. The minimal energy configuration for the SSEP and SIP is sketched in the sub-figures of (a) and (b) correspondingly. For the SSEP, the particles pack tightly, leading to $U_{top} = U(\overline{\rho}_0)$. For the SIP, all the particle can accumulate on the potential minimum at $x=0$, leading to $U_{top}=0$. The $g$ function is tested numerically, as a function of $\overline{\rho}_0$, for the SSEP (a) and the SIP (b) with three potentials $U(x)$: linear  $\Delta U x$, harmonic $\Delta U x^2$ and cosine  $ \frac{1}{2}\Delta U (1-\cos(\pi x  ))$. We confirm numerically, for the SSEP $g=1-U(\overline{\rho}_0)/\Delta U$ (left panel) and for the SIP $g=1$ (right panel), as predicted by Eq.~\eqref{gen-AL}. Both predictions are well fitted up to the expected numerical errors. We refer to \cite{SI} for more details. 
}
\label{fig:generalized AL plots}
\end{center}
\end{figure*}

In this letter, our aim is to delve deeper into the AL for interacting systems. However, before doing that, it is insightful to derive the AL for $M$ non-interacting particles. Notably, the AL can be computed from the large time survival probability which can be derived by mapping the problem to an effective Schr\"{o}dinger equation with absorbing boundaries -- see the textbooks \cite{gardiner1985handbook,risken1996fokker} for this standard procedure. Technically, the survival probability $\mathcal{S}(t) \asymp \exp \left[ -t \,  \Phi \right]$ \cite{gardiner1985handbook}, \footnote{Assuming $t \gg L^2/D_0 $ the diffusion time. For the short time limit, see \cite{agranov2018narrow}.}, where $\Phi = D_0 M \Lambda$ and $\Lambda$ is the ground state energy of the Fokker-Planck Hamiltonian $\hat{H}_{FP} =-D_0 \partial_{xx} + \frac{D_0}{4} (\partial_x U /D_0 )^2  - \frac{1}{2} \partial_{xx}U$ \cite{SI,sabhapandit2020freezing}. Evidently, the AL universality still prevails for many non-interacting particles. What happens to this universality when we incorporate many-body interactions? Here, we try to address this important question.

Consider an extended 1D system of interacting diffusing particles in a monotonically increasing potential $U(x)$ with initial mean density $\overline{\rho}_0 = M/L$, where $M$ is the number of particles and $L$ is the domain size. We show, in this letter, that the mean time it takes for a particle to escape from the potential is given by the following generalized AL
\begin{equation}
    \Phi \asymp \rm{e}^{-\Delta U g(\overline{\rho}_0 )  / D_0  } ,
    \label{gen-AL}
\end{equation}
where $g$ depends both on the 
density $\overline{\rho}_0$ and on the inter-particle interactions. By minimizing the particle configuration energy, one can show  $ g = 1-  U_{top}/\Delta U $ where $ U_{top}$ is the highest energy a particle can attain in the minimum energy configuration. The salient point here is that the exponential form of the AL is still preserved. However, the generalized AL crucially depends on the exact nature of the inter-particle interactions as well the density.

To illustrate the generalized AL, consider the paradigm of simple symmetric exclusion process (SSEP) where the inter-particle interaction is a hard-core exclusion \cite{mallick2015exclusion}. Minimizing the particle configuration energy implies that they are tightly packed around the potential minimum, resulting in $g<1$ [see Fig.~\ref{fig:generalized AL plots}(a), details later]. Therefore, the repulsive interaction imposes a shorter mean escape time by (\ref{gen-AL}). Eq.~\eqref{gen-AL} is the central result of this letter and we will prove it for a class of diffusive interacting systems, within the framework of the macroscopic fluctuation theory (MFT) \cite{bertini2015macroscopic}.

\emph{Macroscopic Fluctuation Theory.---}
Over the last two decades, the Macroscopic Fluctuation Theory (MFT) has been instrumental to understand nonequilibrium fluctuations in diffusive systems \cite{derrida2001free,bertini2002macroscopic,appert2008universal,shpielberg2018universality,tailleur2008mapping,derrida2019large,hurtado2014thermodynamics,mallick2022exact,derrida2007non,bertini2006non,imparato2009equilibriumlike,baek2017dynamical,bettelheim2022inverse,lazarescu2015physicist,tizon2017order,agranov2023macroscopic}. Quite recently, the MFT was also used to capture the survival probability of interacting diffusive particles from a domain 
\cite{agranov2018narrow}. Here, we extend this formalism in the presence of a potential which naturally allows us to compute the survival probability and study the generalized AL.

To set the stage, let us consider a $1D$ system of size $L$. The system is occupied with interacting diffusive particles that satisfy the continuity equation $\partial_s \rho = -\partial_x j$  with the density and current density $\rho(x,s)$ and $j(x,s)$ respectively. Here, $x \in \left[0,1\right]$ and $s \in \left[0,t \right]$ are diffusively rescaled \cite{derrida2007non}.  
The fundamental formula of the MFT asserts that the path probability is 
\begin{eqnarray}
\label{eq:fundamental MFT}
    \text{Prob}\left[ \lbrace \rho , j  \rbrace \right] &\asymp& \exp{  \left(  - \frac{L}{4D_0} \intop^1 _0  dx \intop^{t} _0 ds \,  \frac{(j-J(\rho))^2}{\chi(\rho)}     \right)   }
    \nonumber  \\ 
    J(\rho)&  = & -D(\rho) \partial_x \rho - \chi(\rho) \partial_x \tilde{U},
\end{eqnarray}
where we have defined $\tilde{U}=U/D_0$ as the rescaled potential.  $D(\rho)$ and $\chi(\rho)$ are the density dependent diffusivity and mobility which encapsulate the diffusive dynamics. In Eq.~\eqref{eq:fundamental MFT}, the continuity equation is implicitly assumed. 

Moreover, we assume that the particles are constrained between one reflecting and one absorbing wall at $x=0$ and $x=1$. Thus, $\mathcal{S}(t)$, the survival probability  of the particles to stay inside the region upto time $t$ can be written as a conditional sum over all the paths Prob$\left[ \{ \rho , j  \} \right] $ that satisfy the following boundary conditions
\begin{equation}
\label{eq: BC MFT}
J(\rho)|_{x=0} =  \rho_{x=1} = 0,
\end{equation}
and the mass conservation in the system $L \int dx \rho(x,t) = M$.

Within the MFT, one would expect $\mathcal{S}(t) \asymp \rm{e}^{-t \, \Phi} $, where computation of $\Phi$ reduces to a minimization problem of finding  an optimal fluctuation $\lbrace \rho,j \rbrace $ that satisfy the above mentioned constraints. Note that the optimal fluctuation governs $\Phi$ due to the large $L$ saddle dominated probability in \eqref{eq:fundamental MFT}, however the minimization problem still remains hard. To address that, we use the additivity principle (AP), that was introduced in 
\cite{bodineau2004current} and proved as a useful tool in evaluating large deviations \cite{bertini2015macroscopic} (also see \cite{shpielberg2016chatelier,shpielberg2017numerical,shpielberg2018universality,appert2008universal}). The AP posits that the optimal fluctuation density is time-independent, reducing the complexity of finding the optimal fluctuations.

For a $1D$ system, the AP assumption $\rho(x,t) = \rho(x) $ implies $j(x,t)= $const due to the continuity equation. However, const$=0$ since the current vanishes at the boundaries. Finding $\Phi$ then reduces to the following minimization problem 
\begin{equation}
\label{eq: Phi q=0 MFT}
    \Phi = \frac{D_0 L}{4} \min_{\rho(x) }\int dx \,  \mathcal{L}(\rho,\partial_x \rho ), ~  \mathcal{L} = \frac{J(\rho)^2 }{\chi} - 4\Lambda (\rho- \overline{\rho}_0), 
\end{equation}
where $\rho(x)$ is subjected to \eqref{eq: BC MFT} and  $\Lambda$ is Lagrange multiplier ensuring the mass conservation $\overline{\rho}_0 = \int dx \rho(x) $. 

The minimization problem in \eqref{eq: Phi q=0 MFT} then boils down to solving an Euler-Lagrange (EL) equation with the Lagrangian $\mathcal{L}$. Using the transformation 
$F(\rho) = \int^{\rho } _0 dz \frac{D(z)}{2\sqrt{\chi(z)}} $ (see \cite{agranov2018narrow}), the resulting EL reads
\begin{equation}
\label{eq: EL interacting F }
    -\partial_{xx} F + \frac{1}{8} (\partial_x \tilde{U})^2 \chi' - \frac{1}{2} \sqrt{\chi} \partial_{xx} \tilde{U} = \Lambda \frac{\sqrt{\chi}}{D}, 
\end{equation} 
where $\chi = \chi(\rho[F])$ and $\chi' =\delta \chi / \delta F $. The resulting boundary conditions inherited from \eqref{eq: BC MFT} are $J_F |_{x=0}= \rho[F]|_{x=1}   =0$ with rescaled current $J_F = \partial_x F+ \frac{1}{2}\chi^{1/2}\partial_x \tilde{U}$.  Thus, the survival probability can be estimated via (\ref{eq: Phi q=0 MFT}) as 
\begin{equation}
\label{eq: Phi with JF}
 \Phi = D_0 L \int dx J^2 _F   , 
\end{equation}
where $F$ is the solution of \eqref{eq: EL interacting F }. 
Before discussing the interactions, it is instructive to re-derive the results for non-interacting particles. Indeed, here $D=1, \chi = \rho$ so that $\rho = F^2$ and \eqref{eq: EL interacting F } recovers $\hat{H}_{FP} F = \Lambda F$, which is essentially the eigen-value problem for the non-interacting system. Skipping details, one can show that $\Phi = D_0 L \overline{\rho}_0 \Lambda $ \cite{SI} which implies that $\Lambda$ is the ground state energy of $\hat{H}_{FP}$ as found earlier.

Furthermore, assuming short range interactions, the limit of a dilute system, i.e. $\overline{\rho}_0\rightarrow 0$, suggests interactions become negligible. Indeed in this limit, $F\rightarrow 0$. Thus, Eq.~\eqref{eq: EL interacting F } can be linearized in $F$, leading to the non-interacting case. Thus for dilute systems, the AL is recovered. In what follows, in order to justify the generalized AL in Eq.~\eqref{gen-AL},
we turn to study two  processes  of interacting systems: the SSEP for repulsive interactions and the  symmetric inclusion process (SIP) for attractive interactions.

\emph{Repulsive interaction.---}The SSEP is a lattice gas model where each lattice site $x$ has an occupancy $n_x=\lbrace 0,1 \rbrace$.
In the SSEP, a particle is allowed to hop to a
nearest neighbour with a fixed rate, provided
the target site is empty. Despite its simplicity, the SSEP has been widely studied in the context of nonequilibrium statistical mechanics and has been used as a paradigmatic model for understanding the behavior of interacting particle systems in many physical and biological systems, e.g. traffic flow, protein synthesis, and gene regulation \cite{mallick2015exclusion}. The SSEP typically demonstrates genuine nonequilibrium behavior, e.g. a non-product steady state measure and long range nonequilibrium correlations \cite{bertini2015macroscopic,derrida2007non}. Furthermore, the SSEP is susceptible to exact solutions; microscopically via the Bethe ansatz and macroscopically through the MFT \cite{appert2008universal,shpielberg2018universality,derrida2019largeI,derrida2019large,mallick2022exact}.

For the SSEP, $D=1,\chi=\rho(1-\rho)$ \cite{derrida2007non}. The transformation to $F$ leads to $\rho = \sin^2 F$. We restrict $F$ to the range $ \left[ 0,\pi/2\right]$ to ensure a positive $\chi^{1/2} = \frac{1}{2}\sin 2F$. While $\chi,\chi'$ in \eqref{eq: EL interacting F } becomes explicit, a solution for arbitrary potential is challenging. Fortunately, an analytical solution can be obtained for the linear potential $\tilde{U}(x) = \Delta \tilde{U} x$.  In that case \eqref{eq: EL interacting F } is simply an autonomous equation. With the transformation $y = \cos 2F$, and by assuming that the density as well as $y$ are  monotonous functions, \eqref{eq: EL interacting F } can be reduced to a first order differential equation \cite{SI}
\begin{eqnarray}
\label{eq: y eq}
    \frac{dy}{dx} &=& \frac{1}{2 }\Delta \tilde{U} \sqrt{1-y^2} \sqrt{1-y^2 + 8 \lambda (  y -y_0)}, 
    \nonumber  \\ 
      &&  y_0 = y|_{x=0}, \quad   y_1 = y|_{x=1}=1, 
\end{eqnarray}
where $\lambda = \Lambda/ \Delta \tilde{U}^2$.  Note that $-1 \leq y_0 \leq 1 $. Here $y_0 \rightarrow 1 (-1)$ implies that the density at the reflecting boundary is approaching zero (unity). 

Eqs. \eqref{eq: Phi with JF} and \eqref{eq: y eq} constitute an implicit solution of the survival probability, denoted by  $\Phi_{\rm SSEP}$, for the SSEP. However, it is worth stressing that one need not explicitly solve $y$ (and thus $F$) in order to obtain $\Phi_{\rm SSEP} $. Skipping details from \cite{SI}, we obtain
\begin{align}
    \Phi_{\rm SSEP} &= \frac{C_0 }{2}\left( y_0 -1 + C_0 - C_2 + 4\lambda  (C_1 - y_0 C_0 ) \right), \nonumber \\
    \text{where~~} C_k &= \int^1 _{y_0} dy \frac{y^k}{\sqrt{1-y^2}\sqrt{1-y^2 + 8\lambda (y-y_0)}},
\end{align}
for $k=0,1,2$. Fortunately, the $C_k$ integrals are analytical and involve elliptic functions. Furthermore,  $\Phi_\text{SSEP}$ is given in terms of ($\lambda,y_0$) which are implicit functions of $\Delta \tilde{U}$ and $\overline{\rho}_0$.

To make the relations explicit, we notice that direct integration of  \eqref{eq: y eq} leads to $C_0 = \Delta \tilde{U}/2$. Also, recalling that $\overline{\rho}_0  = \int dx \sin^2 F  $ implies that $2\overline{\rho}_0   = 1- C_1/C_0$. 
The regime of large $\Delta \tilde{U}$ is supported by $0<1+y_0 \ll \lambda \ll 1 $. In these limits and to leading order $\lambda = 2 \rm{e}^{-\Delta \tilde{U} (1-\overline{\rho}_0)} $ and $ (1+y_0) =  \rm{e}^{-\Delta \tilde{U}  \overline{\rho}_0 }$. Finally, we have
\begin{align}
 \Phi_{\rm SSEP} = D_0 L A \rm{e}^{-\Delta \tilde{U} (1-\overline{\rho}_0)},   
 \label{eq: Phi-SSEP}
\end{align}
where $A $ is a polynomial function of $\overline{\rho}_0$ and $\Delta \tilde{U}$ \cite{SI}.

At this point, we connect the above with our announced result \eqref{gen-AL}. Eq.~\eqref{eq: Phi-SSEP} implies $g = 1- \overline{\rho}_0$ for the linear potential. In the exclusion process, the system's energy is minimized if the particles are ordered as close to the potential minimum as possible. Consequently, the minimum energy occurs when the system is fully occupied between $x=0$ and $x=\overline{\rho}_0$, and therefore $U_{top} = U(x=\overline{\rho}_0)$ (see Fig.~\ref{fig:generalized AL plots}).

Although solving Eq.~\eqref{eq: EL interacting F } is hard for generic monotonous potentials, the same physical arguments can be given to infer the generalized AL. Indeed, the minimum energy attained by tight packing the exclusion particles in monotonous potentials, lead to $U_{top} = U(x= \overline{\rho}_0)$. To demonstrate this behavior, we employ a standard shooting method algorithm to solve the eigen-problem \eqref{eq: EL interacting F } for various potentials (see \cite{SI} for more details). The numerical results provide an excellent agreement with Eq.~\eqref{gen-AL} thus validating the generalized AL for arbitrary monotonous potentials.

\textit{Attractive interaction.---}We now turn our attention to the Symmetric Inclusion Process (SIP) where the interaction among the particles is attractive. In the SIP, the jump rate of a particle from site $x$ with occupancy $n_x $ to a neighboring site $y$ with occupancy $n_y$ is $n_x (1+n_y)$. Thus, multiple particles can occupy a given site and the dynamics makes particle clustering favorable 
\cite{grosskinsky2011condensation,grosskinsky2013dynamics,reuveni2012asymmetric}.

For the SIP, one finds  $D=1,\chi= \rho (1+\rho)$. The $F$-transformation implies $\rho = \sinh^2 F $. Notice that unlike the SSEP, $F$ is unbounded for the SIP. Similarly to the SSEP analysis, one can explicitly write the Euler-Lagrange equation \eqref{eq: EL interacting F }, but an explicit solution is again challenging for an arbitrary potential. However, some analytical progress can be made for the linear potential which we discuss in the following.

In this case, Eq.~\eqref{eq: EL interacting F } becomes an autonomous equation just like for the SSEP. Using the transformation  $y = \cosh 2F$ and assuming the the monotonicity of $F$ leads us to \cite{SI}
\begin{equation}
\label{eq: SIP y equation }
    \frac{dy}{dx} = -\frac{1}{2}\Delta \tilde{U } \sqrt{y^2-1}\sqrt{y^2 -1 +8 \lambda (y_0 - y)},
\end{equation}
where we note that $1 \leq y \leq y_0 $ and $y_0 = y|_{x=0}$ may be unbounded. Similarly, to the SSEP, one can express the survival probability $\Phi_\text{SIP}$ for the SIP as \cite{SI}
\begin{equation}
\label{eq: PhiSIP}
    \Phi_\text{SIP} = \frac{C_0}{2} \left( 
1-y_0 + C_2 - C_0  + 4\lambda (y_0 C_0 - C_1) 
    \right) .
\end{equation}
 Here, we have re-defined the $C_k$ integrals for the SIP
\begin{equation}
    C_k = \int^{y_0} _{1}  \frac{y^k dy}{\sqrt{y^2-1 }\sqrt{y^2-1+8\lambda(y_0 -y)}} ,
\end{equation}
which are analytic. Integrating \eqref{eq: SIP y equation } over the whole range implies $C_0 = \Delta \tilde{U}/2$. Then, using $\overline{\rho}_0 = \int dx \sinh^2 F$ we find  $C_1-C_0 = \Delta \tilde{U} \overline{\rho}_0$. 
 The large $\Delta \tilde{U}$ regime occurs for $  \lambda \ll \mu = y_0 \lambda \ll  1$.  Then, following \cite{SI} we find  $\lambda = \rm{e}^{-\Delta \tilde{U}(1+\overline{\rho}_0) } $ and $\mu  =2 \rm{e}^{-\Delta \tilde{U} }$ to leading order. This results in 
 \begin{align}
\Phi_\text{SIP} = D_0 L A \rm{e}^{-\Delta \tilde{U}},   
\label{survival-SIP}
 \end{align}
where $A$ once again is a polynomial function of $\overline{\rho}_0, \Delta \tilde{U}$ \cite{SI}. 
 In the non-interacting limit $\overline{\rho}_0 \rightarrow 0$ with comparable   $\mu$ and $\lambda$ and thus, we recover the AL $\Phi_\text{SIP}\propto  \rm{e}^{-\Delta \tilde{U}}$ \cite{SI}.

Eq.~\eqref{survival-SIP} remarkably coincides with the single particle AL. This indeed stems from the underlying physics of the SIP, where the attractive interactions imply that the minimum energy configuration is attained when all the particles bunch at the potential minimum. Comparing $\Phi_\text{SIP} $ with Eq.~\eqref{gen-AL} results in $g=1$. In fact, this holds for any monotonous potential. To demonstrate this fact, we employ our shooting algorithm to solve \eqref{eq: EL interacting F } for different potentials. The numerical results are illustrated in Fig.~\ref{fig:generalized AL plots} validating the AL for SIP regardless of the particle density $\overline{\rho}_0$. We should stress that the simplicity of Eq.~\eqref{survival-SIP} can be deceptive since there is no \textit{simple} or \textit{trivial} way to derive it.

\textit{Discussion.---} Understanding escape times in  a thermal activation process while accounting for many-body interactions is at the heart of this letter.  
To this end, we have studied the survival probability of interacting particles with diffusive dynamics in a potential trap. In the limit of weak thermal fluctuations $\Delta U \gg D_0 $, we show, using  
the seminal macroscopic fluctuation theory, that the Arrhenius law can be generalized (Eq.~\eqref{gen-AL}) with $\Delta U g$ -- an interaction-dependent effective activation energy. We demonstrate our result for two distinct interaction classes. Particles which repel themselves reveal a decrease in $g$ leading to a facilitated escape from the potential. On the other hand, for purely attractive particles barring condensation, we reach quite surprisingly and non-trivially to an effective single particle problem. It should be stressed that albeit its technical simplicity and intuitively appealing 
physical interpretations, any attempts to generalize the insight gained from single particle AL to many particles with arbitrary interactions remains exorbitantly challenging. Furthermore, the generalized AL holds in store quite a few surprises.

Naively, one expects that the generalized AL would be more restrictive from an experimental stand point. However, Eq.~\eqref{gen-AL} suggests the opposite with two immediate interesting directions.  First, consider e.g., single file colloids in a potential trap, represented by exclusion particles.  By controlling the particle density in the trap, one can infer from $\Phi$ the exact shape of the potential  \cite{bryan2022inferring,dudko2006intrinsic,gieseler2021optical}. Second, experimental control of the potential affecting the colloids as in \cite{thorneywork2020direct} together with Eq. \eqref{gen-AL}, allows to infer the particle density in the system.

Beyond these applications, this work opens new avenues of research. First, it will be important to extend Eq. \eqref{gen-AL} to include arbitrary potential landscape, possibly also to higher dimensions \cite{perez2016weak,akkermans2013universal}. Additionally, it is interesting to consider interactions that cannot be classified as purely attractive/repulsive e.g., the Katz-Leibowitz-Spohn models \cite{katz1983phase,shpielberg2017geometrical}.  A yet unexplored path is to consider the AL for multi-species particles, where $D,\chi$ become matrices \cite{bodineau2011phase}. Exploring how inter-species interactions can serve to speed up/ slow down the activation time would be illuminating and potentially useful both in theory and experiments involving transport of particles or ions through channels.

\emph{Acknowledgements.---}
A.P. gratefully acknowledges research support from the Department of Science and Technology, India, SERB Start-up Research Grant Number SRG/2022/000080 and the Department of Atomic Energy, India. O.S. thanks Tridib Sadhu and Baruch Meerson for fruitful discussions. O.S. also acknowledges the support of the Erwin Schr\"{o}dinger International Institute for Mathematics and Physics during the thematic program DPS22, where the discussions took place.

\bibliography{rrwr}

\end{document}